\documentclass{elsart}

\usepackage{amssymb,amsfonts,graphicx}

\begin{document}

\begin{frontmatter}

\title{Dynamics of the return distribution in the Korean financial market}

\author{Jae-Suk Yang\corauthref{cor1}},
\ead{yang@kaist.ac.kr}\corauth[cor1]{Corresponding author. Fax:
+82-42-869-2510.}\author{Seungbyung Chae, }
\author{Woo-Sung Jung, }
\author{Hie-Tae Moon}

\address{Department of Physics, Korea Advanced Institute of Science and Technology,
Daejeon 305-701, South Korea}

\begin{abstract}

In this paper, we studied the dynamics of the log-return
distribution of the Korean Composition Stock Price Index (KOSPI)
from 1992 to 2004. Based on the microscopic spin model, we
found that while the index during the late 1990s showed a power-law
distribution, the distribution in the early 2000s was exponential.
This change in distribution shape was caused by the
duration and velocity, among other parameters, of the information that
flowed into the market.

\end{abstract}

\begin{keyword}
Econophysics \sep Emerging market \sep Log return \sep Power law
distribution \sep Exponential distribution
\\

\PACS 89.65.Gh \sep 89.75.Fb \sep 89.75.Hc
\end{keyword}

\end{frontmatter}


\section{Introduction}

Interdisciplinary research is now routinely carried out, with
econophysics being one of the most active interdisciplinary fields
\cite{stanley,mantegna,mandelbrot2,jung2,jung3}. Many research
papers on mature markets have already been published. However,
since emerging markets show different characteristics to those of
mature markets, they represent an active field for
econophysicists. The Korean market, one of the foremost emerging
markets, has already been studied by physicists
\cite{jung2,jung3}. We concentrate on the particular properties of
the Korean market through the return distribution.

It is broadly assumed that the distribution of price changes takes the
form of a Gaussian distribution, and that all information is applied to
the market immediately by the efficient market hypothesis (EMH)
\cite{cootner}. Using the EMH, the trading profit with arbitrage cannot be
obtained from the superiority of information. The price changes in an
efficient market cannot be predicted and change randomly. This is suited to
classical economics theory. However, experimental proofs reveal that Gaussian
distributions of price changes do not exist in real markets \cite{mandelbrot1,fama}.

Mandelbrot determined empirically that the tail part of the
distribution is wider and the center of the distribution is
sharper and higher than a Gaussian distribution by examining price
changes of cotton; this distribution of price changes is termed
the L\'evy stable distribution \cite{mandelbrot1}. Fama also found
a L\'evy stable distribution for the New York Stock Exchange
(NYSE) \cite{fama}. After Mandelbrot's study, the distribution of
price changes was identified as non-Gaussian by many researchers
\cite{stanley2,mccauley,silva,vicente}. It was reported that
distributions in mature markets have a power-law tail, while those
in emerging markets have an exponential tail \cite{matal,miranda}.

Silva \emph{et al.} \cite{silva} and Vicente \emph{et al.}
\cite{vicente} reported that the distributions of price changes
vary with time lag. Price changes have a power-law distribution
for a short time lag, and an exponential distribution when the
time lag is long. Moreover, for a very long time lag, the
distribution becomes Gaussian. This transition problem was solved
analytically by Heston using a stochastic model \cite{heston}.

Financial markets are adaptive evolving systems, so the
distributions of price changes are also different for various
periods and countries. Especially, the Korean stock market has
different properties compared with other countries, and the
distribution of price changes is not stable and changes with time.

In this paper, we study the characteristics of the Korean stock
market using probability distribution functions (PDFs) of the
Korean Composition Stock Price Index (KOSPI) and investigate why
phenomena different to other countries occur. We also carry out a
simulation using the microscopic spin model to explain these
differences.

\section{Empirical data and analysis}

We use the KOSPI data for the period from 1992 to 2004, and
observe the PDFs of the KOSPI price changes for a time window
of 1 year. We use only intra-day returns to exclude discontinuous
jumps between the previous day's close and the next day's open
price due to overnight effects. The price change log return
is defined by:

\begin{equation}
S(t)\equiv \ln Y(t+\Delta t)-\ln Y(t),
\end{equation}

where $Y(t)$ is the price at time $t$ and $\Delta$$t$ is the time lag.

\begin{figure}
\includegraphics[angle=0,width=0.5\textwidth]{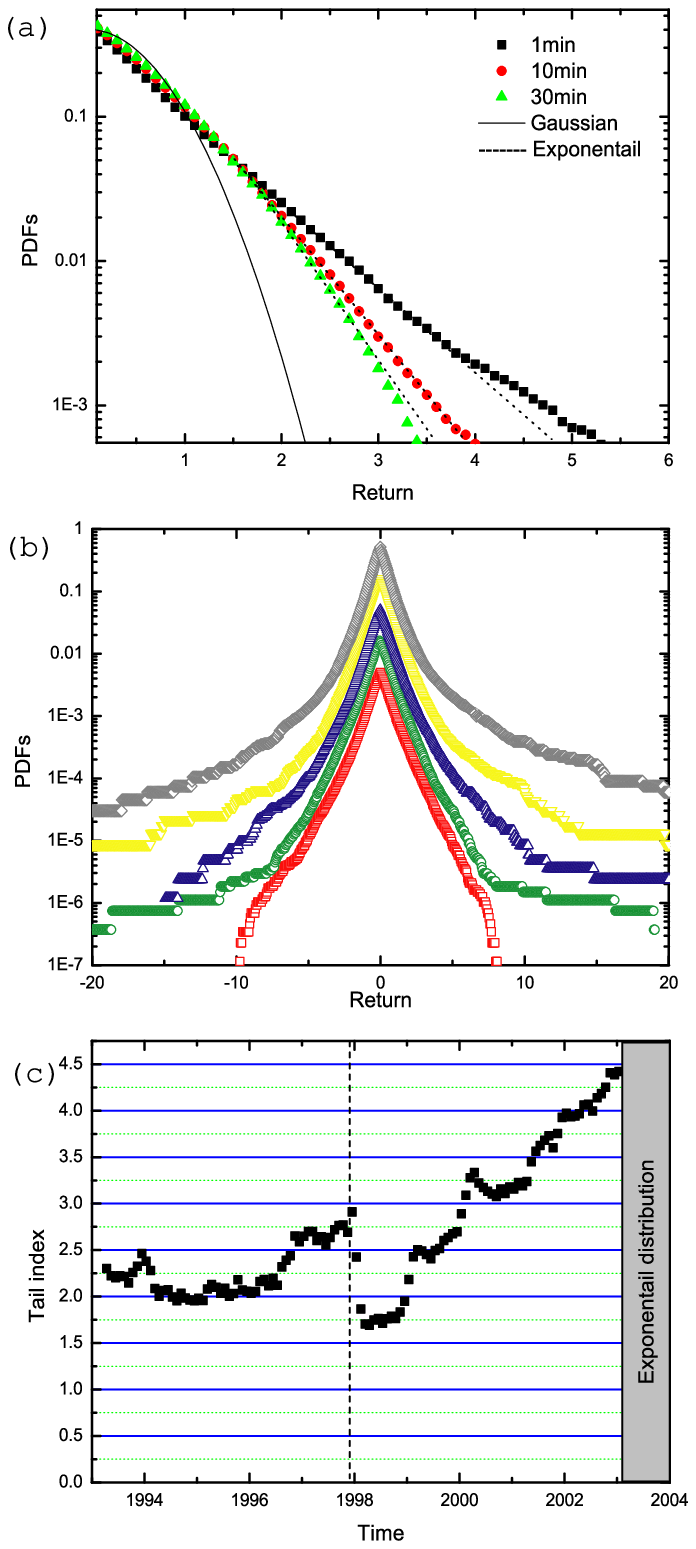}
\caption{(a) Probability distribution functions (PDFs) of log
return for the KOSPI. Rectangle($\blacksquare$) represents 1-min
time lag, circle($\bullet$) 10-min, and triangle($\blacktriangle$)
30-min, respectively. (b) PDFs of log return for the KOSPI from
1998 to 2002: $\Diamond$ 1998, $\bigtriangledown$ 1999,
$\bigtriangleup$ 2000, $\bigcirc$ 2001, and $\Box$ 2002. (c)
Evolution of the KOSPI tail index. The tail first increases and
then changes to exponential. The dashed line represents the 1997
Asian financial crisis.} \label{Fig1}
\end{figure}

Fig. \ref{Fig1}a shows the log return distribution for the KOSPI.
The distribution for an 1-min time lag represents a power-law
distribution, that for 10 min is exponential, and that for 30 min
is also exponential but close to Gaussian. These results are in
accordance with previous reports \cite{silva,vicente}.

In the Korean stock market, the log return distribution for an
1-min time lag shows some peculiar phenomena. Fig. \ref{Fig1}b
shows the PDFs of the KOSPI log return from 1998 to 2002, while
Fig. \ref{Fig1}c is a graph of the tail index, power law exponent
of tail part of the log return distribution, as a function of time
from 1993 to 2004. The shape of the distribution in 1998
($\Diamond$) in Fig. \ref{Fig1}b is close to a L\'evy distribution
and the tail part shows a power-law distribution. However, the
tail index of the PDFs increases over the years and the shape of
tail part changes to exponential with increasing time. This
phenomenon can be confirmed in Fig. \ref{Fig1}c. As well the tail
index in the early 1990s is approximately 2.0, increases from the
mid-1990s to the early 2000s, finally changes to an exponential
tail. Although a discontinuity in the increasing trend occurred
during the 1997 Asian financial crisis, the tail index continued
to increase thereafter.

\begin{figure}
\includegraphics[angle=0,width=1.0\textwidth]{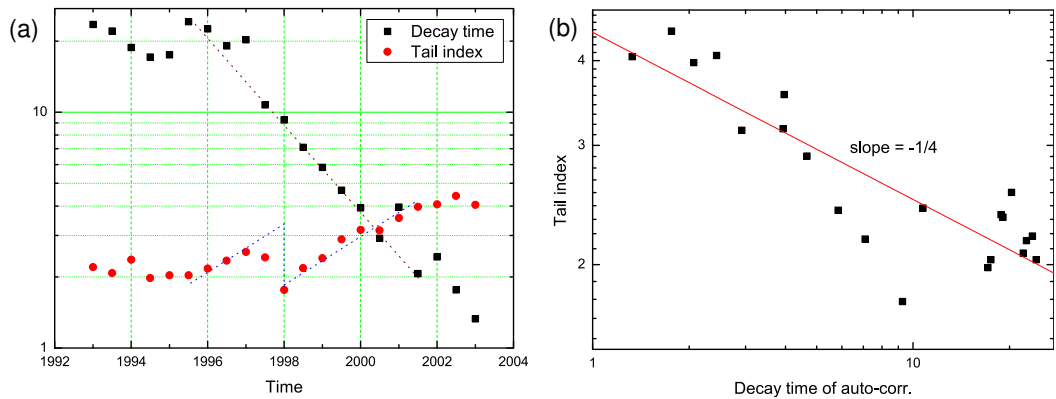}
\caption{(a) Evolution of the decay time for the autocorrelation function (ACF)
and the tail index. (b) Relation between decay time and tail index for the ACF.
} \label{Fig2}
\end{figure}

In Fig. \ref{Fig2}a, the decay time for the autocorrelation
function of log return is continuously decreasing, while the tail
index abruptly varied around the time of the 1997 Asian financial
crisis. This suggests that the decay time is related to the
increasing trend of the tail index, regardless of the Asian
financial crisis. Thus, we investigated the relation between the
decay time for the autocorrelation function of log return and the
tail index of the log return distribution to identify why this
phenomenon is happened. Fig. \ref{Fig2}b shows the relation
between the tail index and the decay time. The decay time of the
autocorrelation function is inversely proportional to the tail
index by a factor of 4.

The decay time of the autocorrelation function decreases as the
log return distribution of the KOSPI changes from a power-law to
an exponential distribution (Fig. \ref{Fig2}). This decrease in
decay time means that the duration of information in the market is
diminished compared with the past, that is, the market is more
rapidly affected by the information flow and the influence of past
information decreases more rapidly when the decay time is longer.
Information and communication technology such as high-speed
internet connections and electronic trading systems were not fully
utilized in the past, so it took a longer time to deliver
information to the market. Moreover, the information flow was
small because social structures in the past were relatively
simple. For this reason, much more information flows into the
market for a specific time interval now compared with the past.
Therefore, the time scale of the past differs from the current
scale. The amount of information and its velocity in reaching the
market for 1 min now may be the same as that for 2 or 3 min or
more in the past.

\section{Model and results}
Eguiluz and Zimmermann \cite{eguiluz}, Krawiecki \emph{et al.}
\cite{krawiecki}, Chowdhury and Stauffer \cite{chowdhury}, and
Cont and Bouchaud \cite{cont} used microscopic models to simulate
the financial market, and these models describe well the
characteristics of the financial market. Krawiecki \emph{et al.}
described the market as a spin model. Agents and traders
comprising the market are represented by spins, and interaction
between agents and external information is represented by fields.

\begin{figure}
\includegraphics[angle=0,width=1.0\textwidth]{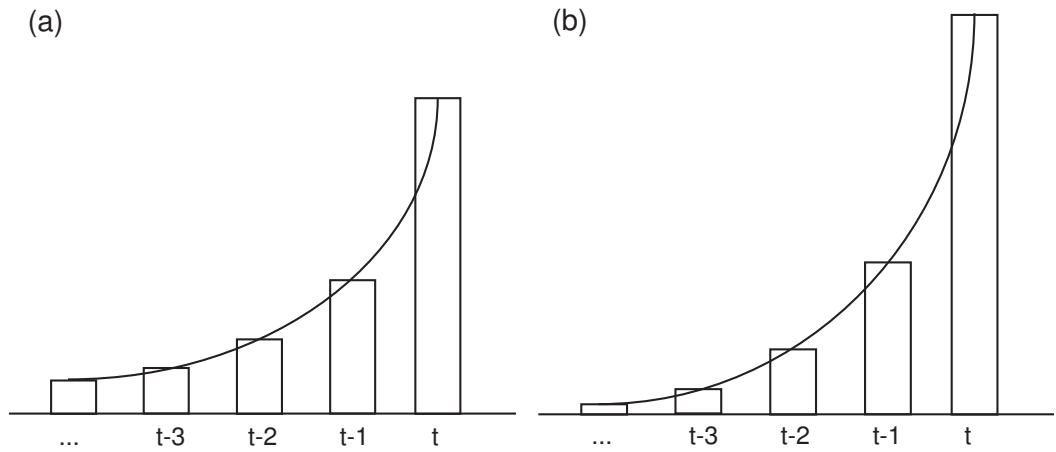}
\caption{Influence of information flow into the market. The
latest information has a strong influence, and influence on the
market diminishes as time passes. (a) If the information flow is small,
the influence of the information slowly decreases. (b) If the information flow
is large, the influence of the information rapidly decreases. The
amount of information used by traders to determine whether to sell or buy is
constant. Thus, the area of these functions has to be constant.}
\label{Fig3}
\end{figure}

We modify the Krawiecki microscopic model of many interacting
agents to simulate the variation of log return distribution for
the Korean stock market. The number of agents is $N$, and we
consider $i=1,2,\dots ,N$ agents with orientations $\sigma_{i} (t)
= \pm 1$, corresponding to the decision to sell ($-1$) or buy
($+1$) stock at discrete time-steps $t$. The orientation of agent
$i$ at the next step, $\sigma_{i}(t+1)$, depends on the local
field:

\begin{equation}
I_{i} (t) = \frac{1}{N} \sum _{j} A_{ij} (t) \sigma_{j} (t) +
h_{i} (t),
\end{equation}

where $A_{ij} (t)$ represent the time-dependent interaction strength among
agents, and $h_{i} (t)$ is an external field reflecting the effect
of the environment. The time-dependent interaction strength among
agents is:

\begin{equation}
A_{ij} (t) = A\xi (t) + a \eta_{ij} (t),
\end{equation}

with $\xi(t)$ and $\eta_{ij} (t)$ determined randomly in every
step. $A$ is an average interaction strength and $a$ is a
deviation of the individual interaction strength. The external
field reflecting the effect of the environment is:

\begin{equation}
h_{i}=h \sum _{k=0}^{\infty} \zeta_{i} (t-k) e^{-k/\tau},
\end{equation}

where $h$ is an information diffusion factor, and $\zeta_{i}(t)$
is an event happening at time $t$ and influencing the $i$-th
agent. $\tau$ is the duration time of the information, which
represents how long the event at time $t$ retains influence on the
opinion of agents on market prices. At every step, agents are
assumed to receive newly generated information. The later the
event, the greater is the influence on agents and the market. The
most recent event has a strong influence on agents, while old
events have a weak influence. Influence on the market is
considered to be reduced exponentially. Moreover, for larger
$\tau$, an event at time $t$ retains a relatively strong influence
on agent opinions and market price changes for a long time. On the
other hand, for shorter $\tau$, the information from an event is
rapidly delivered to agents and applied to market prices
immediately, so the information quickly vanishes from the market
(Fig \ref{Fig3}). If we assume that the information flow is the
same ($=1$) whether $\tau$ is long or short, then $h$ is equal to
1/$\tau$.

From the local field determined as above, agent opinions in the
next step are determined by:

\begin{equation}
\sigma_{i} (t+1) =\left\{ \begin{array}{ll} +1 & \mbox{with probability
$p$}
\\ -1 & \mbox{with probability $1-p$} \end{array} \right..
\end{equation}

where $p=1/(1+exp\{-2I_{i}(t)\})$. In this model, price changes are:

\begin{equation}
x(t) = \frac{1}{N} \sum \sigma_{i}(t).
\end{equation}

We simulate a stock market with 1000 agents, and the values
$\xi(t)$, $\eta(t)$, and $\zeta(t)$ are generated randomly within
the range $[-1, 1]$.

\begin{figure}
\includegraphics[angle=0,width=1.0\textwidth]{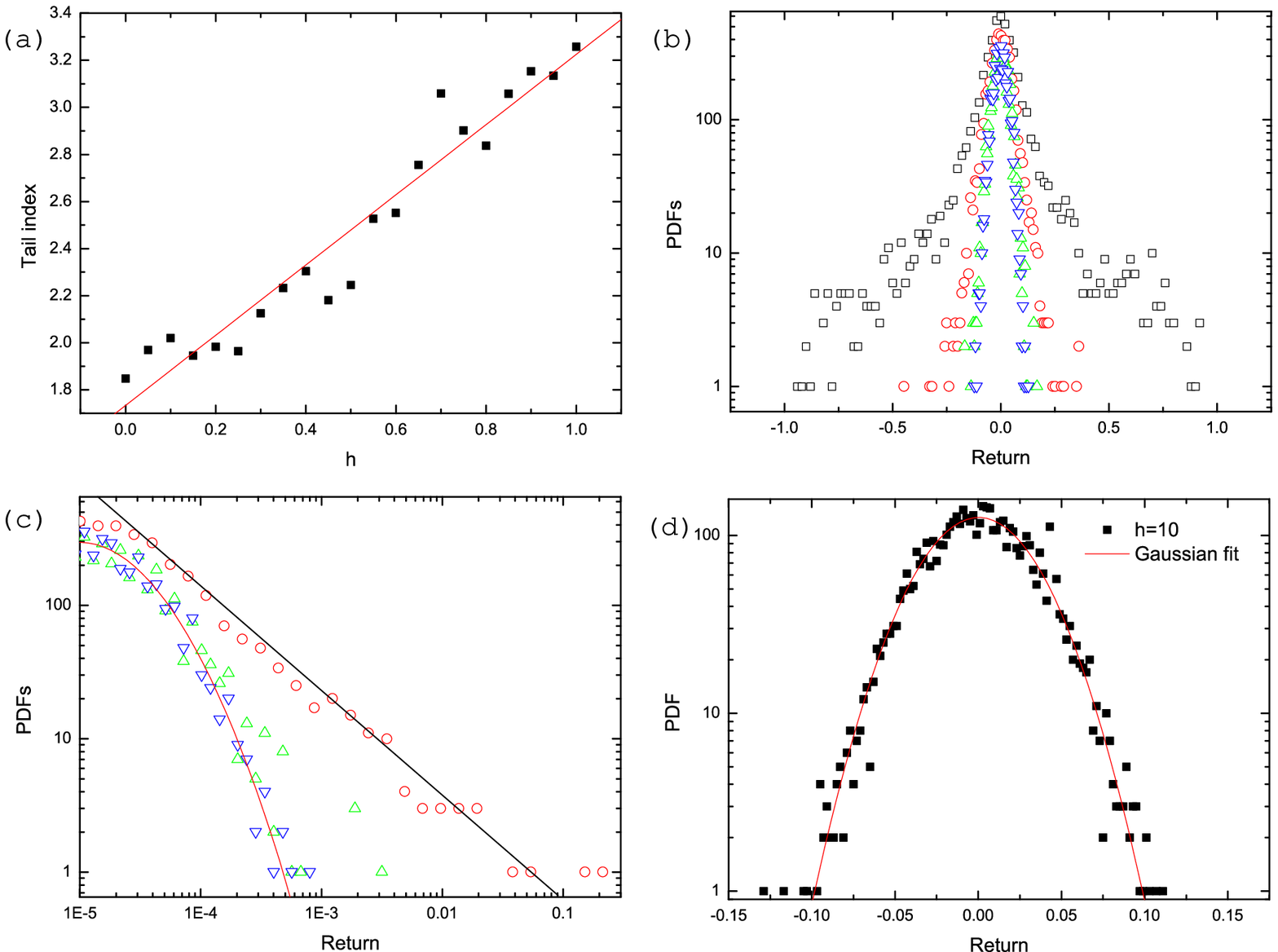}
\caption{(a) Relation between $h(=1/\tau)$ and the tail index of
PDFs. (b) Semi-log plot for the PDFs of price changes as a
function of $h$. $\Box,\ h=0$; $\bigcirc,\ h=1$; $\bigtriangleup,\
h=2$; and $\bigtriangledown,\ h=3$. (c) Log-log plot for the PDFs.
The straight line is guide line for power law distribution and the
curved line is for exponential distribution. Symbols are as for
(b). (d) PDF of price changes when $h$ is very large ($h=10$).}
\label{Fig4}
\end{figure}

Fig. \ref{Fig4}a shows the relation between the diffusion factor,
$h$, and the tail index of the PDFs. The tail index is directly
proportional to $h$. This is similar to the trend line (solid
line) in Fig. \ref{Fig1}c. Fig. \ref{Fig4}b and c show the PDFs of
log return for various $h$ values. Similar to the experimental
result, a power-law tail is evident for small $h$ (large $\tau$),
and an exponential tail for large $h$ (small $\tau$). For very
large $h$, the distribution is close to Gaussian (Fig.
\ref{Fig4}d)

The time lag, $\Delta t$, can be determined for 1 min, 10 min, 1
h, 1 day, etc., when price changes are calculated. However, time
is not uniform, because the volume, the number of contracts, and
the information flow into the market are not homogeneous. In other
words, 1 min of today may not be same as 1 min of tomorrow. Thus,
it is necessary to study the time uniformity.

\section{Conclusions}
The distribution of the KOSPI log return has recently taken the form of an
exponential, while it showed a power-law tail in the 1990s. Moreover, the
decay time of the autocorrelation function is continually decreasing.
Thus, the duration of information received by agents is also decreasing as
the amount of information increases. According to the EMH, the distribution
of log return becomes Gaussian when the velocity of
information flow is very fast, and all information received immediately
affects the opinion of agents in the market.
We could identify and confirm a relationship between the distribution
of price changes, the velocity of information flow, and the duration of
the influence of information for a time series of the Korean
stock market. A similar phenomenon occurred in Japan around
1990, as identified from daily data \cite{kaizoji}. However, in mature markets,
including the NYSE, the tail index is not increasing or changing in shape.
The reason for this is the robustness and maturity of the
market. Mature markets are solid enough to endure external shocks,
while emerging markets are susceptible to shocks and environmental
changes. Modeling of the robustness and maturity of the market is
planned as further work in the near future.

\begin{ack}
We wish to thank H. Jeong and O. Kwon for helpful discussions and
supports.
\end{ack}

\newpage

\begin{thebibliography}{00}

\bibitem{stanley} H.E. Stanley, L.A.N. Amaral, D. Canning, P. Gopikrishnan,
Y. Lee, Y. Liu, Physica A 269 (1999) 156.

\bibitem{mantegna} R.N. Mantegna, H.E. Stanley, An Introduction to Econophysics:
Correlations and Complexity in Finance, Cambridge University
Press, 2000.

\bibitem{mandelbrot2} B.B. Mandelbrot, Fractals and Scaling in Finance:
Discontinuity, Concentration, Risk, Springer, New York, 1997.

\bibitem{jung2} W.-S. Jung, S. Chae, J.-S. Yang, H.-T. Moon, Physica A, in press (Physics/0504009, 2006).

\bibitem{jung3} W.-S. Jung, O. Kwon, J.-S. Yang, H.-T. Moon, J. Korean Phys. Soc.,
in press (Physics/0509090, 2006).

\bibitem{cootner} P.H. Cootner, Ed.,
The Random Character of Stock Market Prices, MIT Press, Cambridge
MA, 1964.

\bibitem{mandelbrot1} B.B. Mandelbrot, J. Bus. 36 (1963) 394.

\bibitem{fama} E.F. Fama, J. Bus. 38 (1965) 34.

\bibitem{stanley2} H.E. Stanley, L.A.N. Amaral, X. Gabaix, P.
Gopikrishnan, V. Plerou, Physica A 299 (2001) 1.

\bibitem{mccauley} J.L. McCauley, G.H. Gunaratne, Physica A 329 (2003) 178.

\bibitem{silva} A.C. Silva, R.E. Prange, V.M. Yakovenko, Physica A 344 (2004) 227.

\bibitem{vicente} R. Vicente, C.M. de Toledo, V.B.P. Leite, N. Caticha, Cond-mat/0402185, 2004.

\bibitem{matal} K. Matal, M. Pal, H. Salunkay, H.E. Stanley,
Europhys. Lett. 66 (2004) 909.

\bibitem{miranda} L.C. Miranda, R. Riera, Physica A 297 (2001) 509.

\bibitem{heston} S.L. Heston, Rev. Financ. Stud. 6 (1993) 327.

\bibitem{eguiluz} V.M. Eguiluz, M. Zimmermann, Phys. Rev. Lett. 85 (2000) 5659.

\bibitem{krawiecki} A. Krawiecki, J.A. Ho\l yst, D. Helbing, Phys. Rev. Lett. 89 (2002) 158701.

\bibitem{chowdhury} D. Chowdhruy, D. Stauffer, Eur. Phys. J. B 8 (1999) 477.

\bibitem{cont} R. Cont, J.-P. Bouchaud, Macroecon. Dyn. 4 (2000) 170.

\bibitem{kaizoji} T. Kaizoji, Physica A 343 (2004) 662.

\end{thebibliography}
\end{document}